\documentclass[10pt,conference]{IEEEtran}

\usepackage{amsmath,amssymb,amsfonts} 
\usepackage{wasysym} 
\usepackage{textcomp}
\usepackage{pifont}

\usepackage[table]{xcolor} 
\usepackage{booktabs}
\usepackage{multirow}
\usepackage{graphicx}
\usepackage{tabularx}
\usepackage{tabulary}
\usepackage{makecell}

\usepackage{algorithm}
\usepackage{algpseudocode}
\usepackage{listings}
\usepackage{xcolor}
\usepackage{amssymb} 

\definecolor{imgGreen}{rgb}{0.0, 0.6, 0.0}
\definecolor{imgRed}{rgb}{1.0, 0.0, 0.0}
\definecolor{imgComment}{rgb}{0.4, 0.5, 0.6}

\lstdefinestyle{imageStyle}{
    language=C,
    basicstyle=\ttfamily\small,
    keywordstyle=\bfseries\color{imgGreen},
    commentstyle=\itshape\color{imgComment},
    stringstyle=\color{black},
    frame=lines,
    rulecolor=\color{black},
    emph={tree, leaf_idx, indices},
    emphstyle=\color{imgRed},
    breaklines=true,
    breakatwhitespace=true,
    postbreak=\mbox{\textcolor{black}{\ensuremath{\hookrightarrow}}\space},
    showspaces=false,
    showstringspaces=false,
    numbers=none,
    captionpos=b
}

\definecolor{mintedGreen}{rgb}{0,0.6,0}
\definecolor{mintedBlue}{rgb}{0.25,0.5,0.5}
\definecolor{mintedRed}{rgb}{0.7,0.15,0.15}

\lstdefinestyle{mintedcpp}{
    language=C++,
    basicstyle=\ttfamily\footnotesize,
    keywordstyle=\bfseries\color{mintedGreen},
    commentstyle=\itshape\color{mintedBlue},
    stringstyle=\color{mintedRed},
    frame=lines,
    rulecolor=\color{black},
    breaklines=true,
    breakatwhitespace=false,
    keepspaces=true, 
    showspaces=false,
    showstringspaces=false,
    showtabs=false,
    tabsize=4,
    captionpos=b,
    morekeywords={asm, volatile, constexpr, __device__, __global__, __shared__, uint32_t, uint8_t, void, template, dim3, prmt, mad, lo, u32, b32, b64, reg, mov}
}

\usepackage{enumitem}
\usepackage{hyperref}
\usepackage{cite}
\usepackage{caption}
\usepackage{xspace}
\usepackage{soul}
\usepackage{comment}
\usepackage{siunitx}

\def\BibTeX{{\rm B\kern-.05em{\sc i\kern-.025em b}\kern-.08em
    T\kern-.1667em\lower.7ex\hbox{E}\kern-.125emX}}

\definecolor{customgreen}{RGB}{130,179,102}  

\title{HERO-Sign: Hierarchical Tuning and Efficient Compiler-Time GPU Optimizations for SPHINCS$^+$ Signature Generation}
\author{
\IEEEauthorblockN{Yaoyun Zhou}
\IEEEauthorblockA{
Department of EECS\\
University of California, Merced\\
 yzhou96@ucmerced.edu}
\and
\IEEEauthorblockN{Qian Wang}
\IEEEauthorblockA{
Department of Electrical Engineering\\
University of California, Merced\\
qianwang@ucmerced.edu}
}

\begin{document}
\maketitle
\thispagestyle{plain}
\pagestyle{plain}

\begin{abstract}

SPHINCS$^+$ is a stateless hash-based signature scheme known for its strong post-quantum security, but it suffers from slow signature speed due to the heavy use of hash operations. The parallel architecture of GPUs offers a potential advantage for accelerating the computation of SPHINCS$^+$ signatures. However, existing GPU-based optimization efforts for SPHINCS$+$ either do not fully exploit the inherent parallelism of its Merkle Tree-based structure, or lack fine-grained, compiler-level customization tailored to its diverse computational kernels. 

This paper proposes HERO-Sign, which adopts hierarchical tuning methodologies and efficient compiler-time GPU optimizations for SPHINCS$^+$. HERO-Sign rethinks the parallelization potential arising from data independence in SPHINCS$^+$' components, including FORS (Forest of Random Subsets), MSS (Merkle Signature
Scheme, and WOTS$^+$ (Winternitz One-Time Signature Plus). First, it introduces a Tree Fusion strategy for FORS, whose structure contains a large number of branches. Our FORS Fusion strategy is supported by an automated Tree Tuning search algorithm, allowing it to adapt and optimize fusion schemes across various GPU platforms. To further enhance performance, HERO-Sign adopts an adaptive compilation strategy that accounts for the varying effectiveness of compiler optimizations across different SPHINCS$^+$ component kernels (\texttt{FORS\_Sign}, \texttt{TREE\_Sign}, \texttt{WOTS$^+$\_Sign}). This strategy automatically selects between PTX and native branches during the compilation phase to maximize efficiency. For multiple batches of message signatures, HERO-Sign focuses on optimizing kernel-level overlapping and employs a Task Graph-based construction strategy to minimize multi-stream idle time and reduce kernel launch overhead. Compared to state-of-the-art GPU implementations, under the SPHINCS$^+$-128f, SPHINCS$^+$-192f and SPHINCS$^+$-256f parameter sets, HERO-Sign demonstrates an enhanced throughput of 1.28$\times$–3.13$\times$, 1.28$\times$–2.92$\times$, and 1.24$\times$–2.60$\times$ on RTX 4090. Similar performance improvements have also been achieved on other architectures, including the A100, H100, and GTX 2080. HERO-Sign also achieves a two-order-of-magnitude reduction in kernel launch latency.


\end{abstract}




\maketitle


\section{Introduction}



Quantum computing has emerged as a transformative technology with profound implications across multiple domains, including cryptography, drug discovery, materials science, and artificial intelligence \cite{Preskill_2018, 8585034, Alan, biamonte2017quantum}. However, with the improvement of quantum error correction techniques, widely used public-key encryption systems such as Rivest-Shamir-Adleman (RSA) and Elliptic Curve Cryptography (ECC) are facing new security challenges. Shor’s algorithm threatens the security foundations of RSA and ECC by enabling efficient solutions to integer factorization and discrete logarithm problems, while Grover’s algorithm presents a moderate risk to symmetric-key systems through quadratic speedups ~\cite{365700,grover1998framework}. The advent of IBM’s 1,121-qubit Condor quantum processor significantly advances the practical feasibility of implementing such quantum cryptanalysis algorithms \cite{ibm_quantum_roadmap}. 

Based on this, NIST launched the Post-Quantum Cryptography (PQC) standardization initiative to evaluate quantum-resistant cryptographic algorithms, culminating with the standardization of lattice-based schemes for key encapsulation (CRYSTALS-Kyber, Hamming Quasi-Cyclic) and digital signatures (CRYSTALS-Dilithium and FALCON), alongside the hash-based signature scheme SPHINCS$^+$ \cite{NIST2025HQC,moody2021nist, 10.1145/3319535.3363229}. Among these algorithms, SPHINCS$^+$ uses more than 100,000 hash computations in Hypertree Structure. Although it provides strong security guarantees based solely on the hardness of hash functions, it suffers from low computational efficiency. For instance, SPHINCS$^+$-128f has larger signature sizes of 17,088 bytes, compared to lattice-based schemes such as FALCON-512 (690 bytes) and Dilithium-2 (2,420 bytes). On the Arm Cortex-A72 platform, SPHINCS$^+$-256f signature generation is approximately 18 times slower than FALCON-512 and 259 times slower than Dilithium-2 \cite{dong2024evaluatingpostquantumcryptographyembedded}. Specifically, in high-throughput applications such as blockchain, authentication, VPNs, and IoT devices, the signatures speed of SPHINCS$^+$ directly impacts the performance of the system.

Prior main optimization works for SPHINCS$^+$ can be categorized into three approaches: FPGA-based custom accelerators: Optimizing latency-critical paths~\cite{amiet2018fpga, berthet2021area}. CPU-level enhancements: Using SIMD extensions, e.g., AVX2/AVX-512/Neon, cryptographic instruction sets (e.g., Intel SHA-NI)~\cite{hanson2022optimization, alter2021optimizing, 11014474, liboqs}. GPU-centric parallelization: Exploiting the algorithm’s inherent parallelism in SPHINCS$^+$ structure~\cite{kim2024parallel, 9095410, 10677363}. GPUs offer thousands of CUDA Cores, a fine-grained SIMT (Single Instruction, Multiple Threads) execution model and hierarchical memory subsystems (Global Memory, Shared memory, and Constant Memory, etc.), which provide a flexible framework for us to optimize SPHINCS$^+$’s compute-intensive operations. 

\begin{figure*}[h]
  \centering\includegraphics[width=1.0\linewidth]{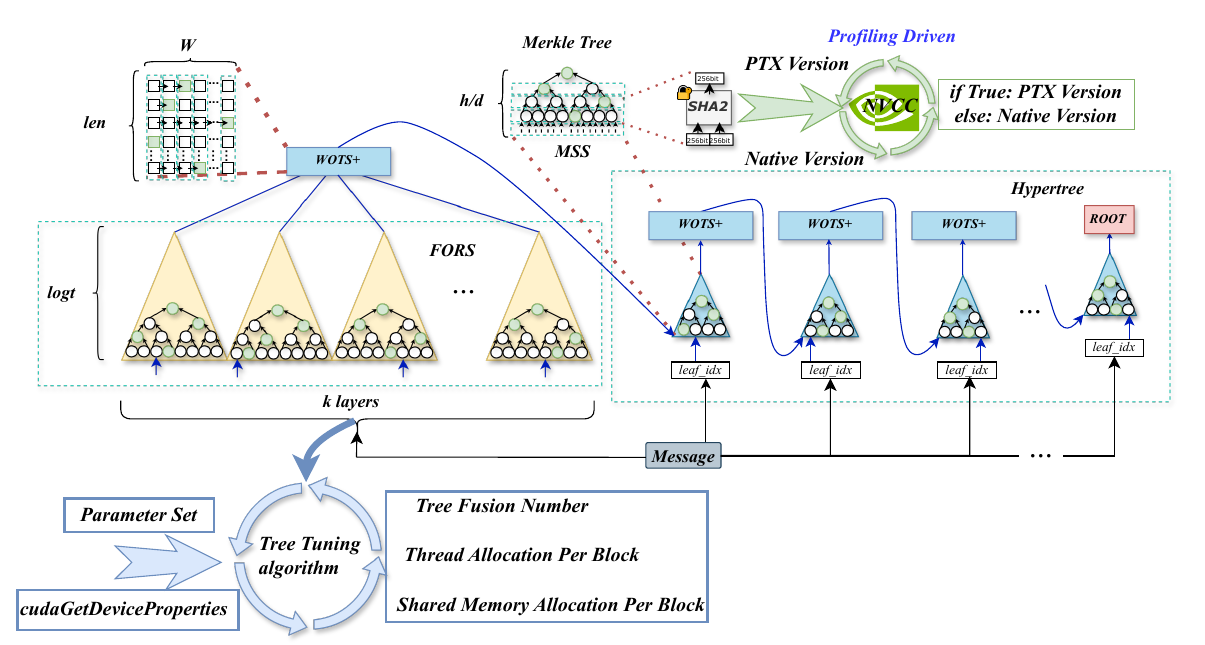}
  \caption{ The overall structure of SPHINCS$^+$, highlighting the key components: FORS, WOTS+, MSS and the resulting Hypertree structure. 
  \textcolor{blue!80!black}{\textrightarrow} represents the flow of signature computation and storage from left to right.   
  \scalebox{1}{\textcolor{customgreen}{\CIRCLE}} \textcolor{customgreen}{\rule{0.6em}{0.6em}} denote the generated signatures. Merkle tree signatures are computed bottom-up and WOTS+ signatures are derived from multiple WOTS+ chains. \textcolor[rgb]{0.125,0.773,0.698}{\dashbox{1}(8,8){}} marks the parallelizable regions of the structure. It further illustrates the workflow the Tree Tuning algorithm and the compile-time execution path selection mechanism.}
  \vspace{-0.2cm}
  \label{fig:sphincs}
\end{figure*}

Different from previous GPU-based works, our work not only introduces a hierarchical optimization framework but also provides a more flexible auto-tuning strategy to adapt to various GPU platforms. We select SHA-256 as the hash function baseline for SPHINCS$^+$ due to its widespread adoption and standardization. Our optimization strategies are algorithm-agnostic and do not depend on specific hash function. Therefore, users can flexibly adopt other standardized hash primitives (e.g,. SHA-512, SHA-3), according to their security and performance requirements. Furthermore, we focus on the \texttt{-f} parameter sets, which feature a larger number of FORS trees and longer WOTS chains than the \texttt{-s} variants, thus presenting greater optimization challenges. 




Our work contributions are as follows:

\begin{itemize}[topsep=2pt, partopsep=2pt]

    \item We propose HERO-Sign, a CUDA-based implementation on GPU, which adopts \underline{H}ierarchical Tuning and \underline{E}fficient Compile\underline{R}-Time \underline{O}ptimizations for SPHINCS$^+$ \underline{Sign}ature Generation. HERO-Sign rethinks the potential of inherent data parallelism in SPHINCS$^+$'s components, where MSS are parallelized based on tree-level parallelism, and WOTS$^+$ is parallelized using chain-level parallelism. 

    \item 
    HERO-Sign leverages shared memory as a bridge to optimize the highly branched structure of FORS by introducing an automated Tree Tuning search algorithm.
    Further, HERO-Sign develops a more general memory padding strategy for mitigating shared memory multi-bank conflicts, supporting single-threaded access patterns of 16-, 24-, and 32-byte widths.
    
    
    \item HERO-Sign allows fine-grained tuning via PTX-based SHA2 implementations. At compile time, each component kernel adaptively selects between native and PTX-optimized branches based on the effectiveness of compiler optimizations. In contrast to runtime branching, which can increase register pressure and lead to a higher memory footprint due to multiple active code paths, our compile-time branching strategy allocates a single, fixed execution path per kernel with only a small increase in compilation overhead.
    \item For multiple batches of message signatures, HERO-Sign employs a Task Graph-based construction strategy, minimizing idle time and reducing kernel launch overhead. HERO-Sign also explores the impact of varying batch sizes on kernel-level overlapping efficiency.
\end{itemize}




\begin{table}[htbp]
\centering
\captionsetup{font=small} 
\caption{The SPHINCS$^+$-f parameter sets. $n$: The bytes of hash outputs, secret keys and public seeds. $h$: The height of the hypertree. $d$: The number of layers in the hypertree. $\log(t)$: The height of each FORS tree. $k$: The number of FORS trees. $w$: The Winternitz parameter used in WOTS$^+$.}
\vspace{1mm}
\begin{tabular}{@{}lcccccc@{}}
\toprule
\textbf{Scheme} & \textbf{$n$} & \textbf{$h$} & \textbf{$d$} & \textbf{$\log(t)$} & \textbf{$k$} & \textbf{$w$} \\
\midrule
SPHINCS$^+$-128f  & 16 & 66 & 22 & 6  & 33 & 16 \\
SPHINCS$^+$-192f  & 24 & 66 & 22 & 8  & 33 & 16 \\
SPHINCS$^+$-256f  & 32 & 68 & 17 & 9  & 35 & 16 \\
\bottomrule
\end{tabular}
\label{tab:params}
\end{table}

\section{Background and Motivation}
\label{sec:bg}
\subsection{SPHINCS$^+$ Signature Scheme}
SPHINCS$^+$ is characterized by a modular design comprising FORS, WOTS$^+$, MSS, and a multi-layer Hypertree structure. Figure~\ref{fig:sphincs} illustrates the overall structure and signature generation flow. In the following sections, we analyze the parallelization in each component of the signature process. Table~\ref{tab:params} presents the specific values for the f parameter.

\subsubsection{Winternitz One-Time Signature Scheme}
The generation of a WOTS$^+$ (Winternitz One-Time Signature Plus) signature involves computing a sequence of hash chains over a message-derived input. The total number of chains, denoted as $\text{WOTS}_{\text{len}}$, is determined by both the chosen Winternitz parameter $w$ and the output length of the underlying hash function. Specifically, $\text{WOTS}_{\text{len}} = \ell_1 + \ell_2$, where $\ell_1 = \left\lceil \frac{8\cdot n}{\log_2 w} \right\rceil$ encodes the message and $\ell_2$ ensures checksum integrity. Each element of the WOTS$^+$  signature is computed as a hash chain of length determined by the base-$w$ message representation. Since each WOTS+ chain operates independently, the signature generation process is highly parallelizable, making it well-suited for acceleration using CUDA multi-threading programming.

\subsubsection{FORS Signature}
The FORS (Forest of Random Subsets) signature scheme constructs a signature by selecting and computing authentication paths in a forest of Merkle trees. The scheme consists of $k$ binary Merkle trees, each with height $\log t$, where $t = 2^{\log t}$ denotes the number of leaves in each tree. To generate a signature, the message is first mapped to $k$ indices, each indicating a leaf in one of the $k$ trees, so that each Merkle Tree can be computed independently. For each index, the corresponding leaf node and its authentication path (comprising $\log t$ sibling nodes) are computed to form the signature. Due to the structure of the Merkle trees, the computation of nodes at each layer in each FORS subtree is independent. This property allows multiple nodes to be computed in parallel using multi-threading.

\subsubsection{Merkle Signature Scheme}
The Merkle Signature Scheme (MSS) is a hash-based digital signature scheme that constructs a single binary Merkle tree to authenticate multiple one-time signatures. At the core of MSS is a Merkle tree built from the public keys of a set of one-time signature (OTS) key pairs, such as WOTS$^+$. The root of this Merkle tree is the public key of MSS. To generate a signature for a message, a one-time key pair is selected, and the message is signed using the corresponding OTS scheme. Along with the OTS signature, an authentication path from the used leaf to the Merkle root is included to prove its validity. This structure enables MSS to generate many secure signatures using a compact public key and the robustness of hash-based cryptography.

\subsubsection{Hypertree Signature Scheme}
The Hypertree signature scheme extends the MSS by organizing multiple MSS trees into a hierarchical structure composed of $d$ layers of subtrees, with a total height $h$. Each layer consists of several Merkle trees of height $\frac{h}{d}$, where the root of each subtree in a lower layer is signed using a one-time signature and incorporated as a leaf node in the Merkle tree of the layer above. The root of the top-level tree serves as the global public key. To generate a signature, the scheme traverses from the bottom layer to the top, producing one-time signatures and corresponding authentication paths at each level. Due to the tree-based structure, both the computation of each MSS subtree and the generation of their internal nodes and authentication paths can be performed independently. This independence allows for effective parallelization of the Hypertree signature generation.

\subsection{Motivation}
\label{sec:mv}
\begin{table}[t]
\centering
\caption{Time Breakdown (ms) for TCAS-SPHINCSp}
\label{tab:time_breakdown}
\resizebox{0.8\linewidth}{!}{
\begin{tabular}{lcccc}
\toprule
\textbf{Time Measure (ms)} & \textbf{FORS} & \textbf{Idle Time} & \textbf{MSS} & \textbf{WOTS+} \\
\midrule
SPHINCS$^+$-128f & 1.89 & 2.27 & 6.57 & 0.93 \\
SPHINCS$^+$-192f & 7.75 & 2.31 & 10.06 & 1.33 \\
SPHINCS$^+$-256f & 13.25 & 2.29 & 26.55 & 1.47 \\
\bottomrule
\end{tabular}
}
\end{table}

For lattice-based PQC algorithms, prior efforts on GPU acceleration of CRYSTALS-Kyber, CRYSTALS-dilithium, and Falcon have been explored~\cite{10.1007/978-3-031-17143-7_25, shen2024high, cmc.2023.033910}. For hash-based schemes like SPHINCS$^+$, Sun et al. \cite{9095410} proposed an initial GPU-based optimization of SPHINCS, the predecessor of SPHINCS$^+$, where the original HORST (Hash to Obtain a Random Subset with Tree) was employed as a few-time signature scheme. This work implemented parallelization strategies for both WOTS+ and HORST over a single Merkle tree. Kim et al. \cite{kim2024parallel} introduced the first GPU-parallel implementation of the Hypertree structure in SPHINCS$^+$. This work extended the single Merkle tree parallelism to support simultaneous computation across multiple Merkle trees, significantly improving throughput. However, their implementation focused primarily on the Hypertree component and supported only single FORS subtree parallelism. Table ~\ref{tab:time_breakdown} presents the detailed time breakdown of their GPU-based SPHINCS$^+$ implementation. We observe that the MSS phase dominates the most computation, followed by FORS, while WOTS$^+$ is relatively lightweight. Meanwhile, a non-negligible amount of idle time is observed across all three parameter sets. CBPSPX introduces a theoretically estimated Thread Utilization Efficiency Index to guide parallel methods selection, but lacks empirical profiling of actual parallel execution efficiency\cite{11068129}.


Furthermore, batch message signatures based on multiple streams has been widely adopted in cryptographic systems, as it enables efficient overlapping of computation and data movement. CUSPX estimates the number of streams by dividing the total number of tasks by the available CUDA cores~\cite{10677363}. While this approach enables effective performance gain, it does not eliminate kernel-level idle time during execution. BatchZK assigns each Merkle tree to a separate CUDA stream and overlaps their execution to improve throughput~\cite{10.1145/3669940.3707270}. While this approach performs well for a small number of Merkle trees, it does not scale efficiently to scenarios like Hypertree in SPHINCS$^+$, which involve up to 35 Merkle trees due to the high latency induced by excessive stream count. Moreover, the degree of overlap between streams is constrained by hardware resources such as register usage, thread allocation, and shared memory, making it difficult to eliminate idle time through stream-level parallelism completely.

To address the above challenges, HERO-Sign rethinks the inherent parallelism within SPHINCS$^+$ and introduces an automated Tree Tuning search algorithm specifically tailored for FORS, which involves a large number of subtrees. For batch message signatures, HERO-Sign explores message-level collaborative parallelism and employs a Task Graph–based kernel construction method to reduce kernel launch overhead and mitigate idle time during concurrent multiple kernels execution. HERO-Sign also prioritizes the use of shared memory for fast memory access whenever possible and selectively tunes instruction choices to align with different component kernels. HERO-Sign is adaptable to performance tuning across diverse GPU architectures.




\section{The Overview of HERO-Sign}
\definecolor{lightblue}{RGB}{200,230,255}


SPHINCS$^+$ is a compute-intensive cryptographic scheme, but the signature generation process of its constituent kernel functions vary significantly. For instance, the generation of leaf nodes in the Hypertree layers heavily relies on repeated calls to the \texttt{wots\_gen\_leaf} function. Taking SPHINCS\textsuperscript{+}-128f as an example, the generation of a single node involves approximately 560 iterations of WOTS$^+$ chain hashing in once \texttt{wots\_gen\_leaf} function call. 
The high register demand per thread increases pressure on the limited register file within each SM, reducing warp occupancy. In contrast, FORS node generation is relatively lightweight per node with only a few hash computations, but due to the large number and increased height of its subtrees, the overall parallel performance depends critically on efficient thread utilization and careful management of shared memory. For WOTS$^+$ signature generation, although it also relies on \texttt{wots\_gen\_leaf}, it is invoked only once per layer, making it more amenable to thread-level parallelization. 


Given these computational characteristics, HERO-Sign follows the approach of Kim et al \cite{kim2024parallel} and decompose the SPHINCS$^+$ signature process into three separate CUDA kernels: \texttt{TREE\_Sign}, \texttt{FORS\_Sign} and \texttt{WOTS+\_Sign}. HERO-Sign integrates parallel Merkle tree processing, a Tree Tuning algorithm for FORS, adaptive compilation branches and Task Graph-based batch message signatures, collectively enhancing GPU resource utilization. In the following Sections A–H, we detail the optimization techniques employed in HERO-Sign.

\label{sec:tf}
\subsection{Multiple Merkle Trees Parallelization} 
\label{sec:MMTP}
\begin{figure}[!ht]
\centering
\begin{lstlisting}[style=imageStyle, caption={SPHINCS$^+$ Signature Generation Snippet}]
hash_message(mhash, &tree, &leaf_idx, sig, pk, 
m, ...);
    ... /* Set WOTS+ Address */
/* Generate the fors's indices array from 
message */
message_to_indices(indices, m);
fors_sign(sig, root, mhash, indices...);
sig += SPX_FORS_BYTES;
for (i = 0; i < SPX_D; i++) {
    merkle_sign(sig, root, ..., leaf_idx);
    ... /* Set Tree Address */
    sig += SPX_WOTS_BYTES + SPX_TREE_HEIGHT * SPX_N;
/* Update the indices for the next layer. */
    leaf_idx = (tree & ((1 << SPX_TREE_HEIGHT) 
             - 1));
    tree = tree >> SPX_TREE_HEIGHT;
}
\end{lstlisting}
\caption{SPHINCS$^+$ Signature Generation Snippet. Both the \texttt{leaf\_idx} and the \texttt{indices} array can be precomputed from a fragment of the input message. The \texttt{indices} array determines the selected signature indices for each FORS subtree, the \texttt{leaf\_idx} specifies the signature index selected within each Hypertree layer.}
\label{fig:leaf}
\end{figure}

As described in the code snippet of Figure~\ref{fig:leaf}, each Merkle tree within the Hypertree structure of SPHINCS\textsuperscript{+} can be computed independently without data dependencies. Similarly, the signature computations for each FORS subtree are also independent. If given sufficient thread blocks and adequate register and shared memory resources, HERO-Sign could parallelize Hypertree and FORS across multiple Merkle trees. In the case of WOTS$^+$, the kernel can be launched once the root nodes of the FORS and Hypertree Merkle trees are computed, enabling further concurrency in the signature generation process. HERO-Sign also applies the \texttt{\_\_launch\_bounds\_\_} qualifier to constrain register usage and ensure reliable kernels' execution under GPU limitations resource.

\subsection{FORS Fusion}
\label{sec:FS}

\subsubsection{Parallelism Feasibility}
Considering SPHINCS$^+$ parameter sets 128f, 192f and 256f, the corresponding Hypertree structures comprise 176, 176, and 272 leaf nodes, respectively. HERO-Sign assigns each leaf node to a dedicated CUDA thread within a block (up to 1024 threads), resulting in shared memory usage of approximately 1KB, 4.125KB, and 8.5KB per block. These allocations remain well within the 48KB per-block static shared memory limit of the RTX 4090, thus enabling fully parallel multiple Merkle tree computations during signature generation. While the FORS contains 2,112, 8,448 and 17,920 leaf nodes respectively, exceeding the 1,024-thread capacity of one block. This results in shared memory consumption of 33 KB, 198 KB and 560 KB, with the latter two configurations surpassing 48 KB shared memory limit per block.

\begin{table}[htbp]
\centering
  \captionsetup{font=small} 
\caption{Warp Occupancy, Theoretical Occupancy, Registers allocated  Per Thread for different component kernels in SPHINCS$^+$-128f in TCAS-SPHINCSp on RTX4090\cite{kim2024parallel}, which are profiled by Nsight System. The \texttt{TREE\_Sign} has low theoretical and practical occupancies with higher register per thread. This high register usage of \texttt{TREE\_Sign} also limits the potential for kernel fusion within SPHINCS\textsuperscript{+}~\cite{7013003,10.1145/1555754.1555775}. We explored with several fusion strategies combining \texttt{TREE\_Sign}, \texttt{WOTS+\_Sign} and \texttt{FORS\_Sign}, but observed performance degradation as a result.}
\begin{tabular}{|p{2.5cm}|c|c|c|} 
\hline
 & \textbf{FORS\_Sign} & \textbf{TREE\_Sign} & \textbf{WOTS+\_Sign} \\
\hline\hline
Warp Occupancy    & 17\%    & 25\%    & 46\%    \\
\hline
Theoretical Occupancy & 66.67\% & 25\%    & 52.08\% \\
\hline
Registers Per Thread  & 64      & 128     & 72      \\
\hline
\end{tabular}
\noindent\footnotesize
\begin{minipage}{0.95\linewidth}
\setlength{\baselineskip}{5pt} 
Occupancy reflects the ratio between the number of active warps and the maximum supported warps per Streaming Multiprocessor (SM).
\end{minipage}
\label{tab:occupancy}
\end{table}

\begin{figure}[h]
  \centering\includegraphics[width=1.0\linewidth]{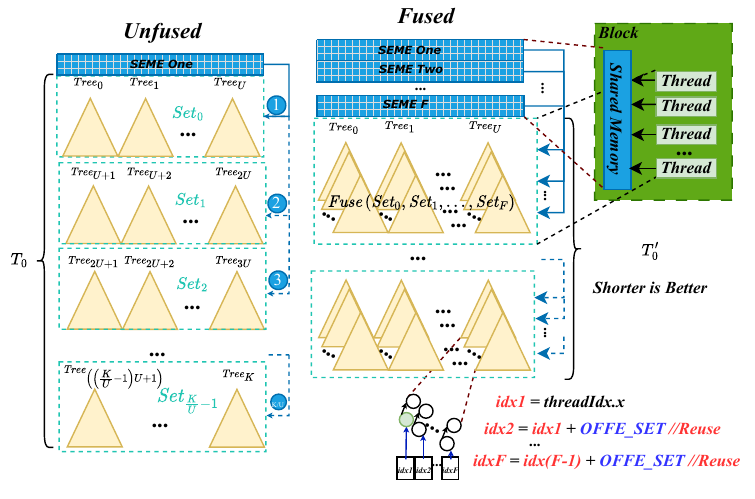}
  \caption{FORS Fusion Process. In Unfused state, for instance, $Set_1$ must wait for $Set_0$ to release \texttt{SEME One}, resulting in a sequential execution pattern in one Block. In Fused state, multiple sets including $Set_0$, $Set_1$, $Set_2$, and up to $Set_F$ are fused to a new $Set$. If sufficient shared memory is available, the value of $F$ could be increased accordingly.}
  \label{fig:tree_fuse}
\end{figure}

\begin{algorithm}[htbp]
\caption{Auto Tree Tuning Algorithm under SEME and Thread Constraints}
\label{alg:adaptive_treefusion_refined}
\begin{algorithmic}[1]
\Require FORS parameters $(k,\;\log_2 t,\;n)$       
\Require $\textsc{SemePerBlock}()$     \Comment{\textcolor{blue}{Dynamic/Static}}              
\Ensure Optimal configuration $(T^\star, F^\star)$

\State $T_{\min} \gets 2^{\log_2 t}$                         \Comment{\textcolor{blue}{Minimum threads per FORS tree}}
\State $T_{\max} \gets 1024$,\quad $S_{\max} \gets \textsc{SemePerBlock}()$
\State $\textit{Cand} \gets \emptyset$

\For{$T_{\text{Set}} = T_{\min}$ \textbf{to} $T_{\max}$ \textbf{step} $T_{\min}$}
    \State $N_{\text{tree}} \gets T_{\text{set}} / T_{\min}$ 
    \State $S_{\text{set}} \gets N_{\text{tree}} \cdot t \cdot n$ 
    \If{$S_{\text{set}} > S_{\max}$}
        \State \textbf{continue}
    \EndIf

   \State $F_{\max} \gets \min\left( \left\lfloor \frac{S_{\max}}{S_{\text{set}}} \right\rfloor,\; \left\lfloor \frac{k}{N_{\text{tree}}} \right\rfloor \right)$

    \For{$F = 1$ \textbf{to} $F_{\max}$}
        \State $T_{\text{used}} \gets T_{\text{set}}$ \Comment{\textcolor{blue}{Threads Fixed per $Set$}}
        \State $S_{\text{used}} \gets F \cdot S_{\text{set}}$
        \Comment{\textcolor{blue}{SEME used after Fusing  $Sets$}}
        \If{$T_{\text{used}} > T_{\max}$ \textbf{or} $S_{\text{used}} > S_{\max}$}
            \State \textbf{continue}
        \EndIf

        \State $U_T \gets T_{\text{used}} / T_{\max}$,\quad $U_S \gets S_{\text{used}} / S_{\max}$
        \If{$U_T = 1$ \textbf{and} $U_S = 1$ \textbf{or} $U_T < \alpha$}  \Comment{\textcolor{blue}{The $\alpha$ value is an optional tune factor and it may vary across different GPU architectures. }}
            \State \textbf{continue}
        \EndIf
     \State $sync \gets \frac{\log_2 t \cdot \left\lceil \frac{k}{N_{\text{tree}}} \right\rceil}{F}$
        \Comment{\textcolor{blue}{Sync points after fusion.}}
        \State $\textit{Cand} \gets \textit{Cand} \cup \{(T_{\text{set}}, F, U_T, U_S, sync)\}$
    \EndFor
\EndFor

\State $(T^\star, F^\star, U_T^\star, U_S^\star) \gets$
\Statex \hspace{\algorithmicindent} $\text{arg\,min}_{(T,F,U_T,U_S,sync)} (sync,\,-U_T,\,-U_S)$ \quad \textbf{over } $\textit{Cand}$
\State \Return $(T^\star,\;F^\star,\;U_T^\star,\;U_S^\star)$
\end{algorithmic}
\end{algorithm}

\subsubsection{FORS Fusion Process}
As shown in Table~\ref{tab:occupancy}, the theoretical occupancy of FORS in the 128f parameter set is 3.89 times higher than the practical occupancy, this suggests notable underutilization of GPU resources. To address this inefficiency, HERO-Sign introduces a FORS Fusion strategy. HERO-Sign first groups multiple Merkle trees that can be processed in parallel within a block into a unit denoted as  $Set$. By leveraging the available shared memory, HERO-Sign allows FORS Fusion operations across consecutive $Sets$, thus enabling parallel execution of multiple $Sets$. This strategy also helps reduce the synchronization overhead typically introduced with multiple independent $Sets$. Figure ~\ref{fig:tree_fuse} illustrates the FORS Fusion process. During the fusion process, HERO-Sign introduces an \texttt{OFFSET} variable to ensure a one-to-one correspondence between the Merkle trees fused from different $Sets$. Since computing \texttt{OFFSET} incurs some overhead due to high branching, HERO-Sign reuses the same variable across fusion process.


\subsubsection{Fusion Trade-offs in FORS}

Determining how many $Sets$ can be fused involves a trade-off between thread allocation and shared memory constraints. HERO-Sign incorporates an offline Auto Tree Tuning algorithm \ref{alg:adaptive_treefusion_refined} to automatically determine the optimal Tree Fusion strategy. The Tree Tuning algorithm automatically queries the maximum shared memory per block available on the target GPU and leverages FORS-specific parameters to compute the optimal fusion strategy. It also identifies a set of near-optimal candidates, among which the final configuration can be selected based on empirical profiling results. The searching process follows a set of prioritized heuristics. First, valid configurations must allocate more threads than the total number of nodes in a single FORS subtree (Line 1), ensuring full coverage. Second, configurations that fully utilize 1024 threads and approach the per-block shared memory limit are excluded (Lines 18-19), as practical results show that such settings tend to increase resource contention and reduce warp occupancy. Third, as the number of fused $Sets$ increases, one synchronization will cover more branches. When multiple near-optimal solutions are found, HERO-Sign prioritizes the configurations that require fewer synchronization points within the candidate set (Lines 21-25). The searching results for the 128f and 192f configurations on RTX4090 are summarized in Table~\ref{tab:fusion_utilization}.
\begin{table}[htbp]
\centering
\captionsetup{font=small} 
\caption{Searching Results of Shared memory utilization(Static), thread utilization, and F (Number of fused $Sets$).}
\renewcommand{\arraystretch}{1.3}
\begin{tabular}{|l|c|c|c|}
\hline
\textbf{Parameter-Sets} & 
\makecell{\textbf{Shared Memory}\\\textbf{Utilization}} & 
\makecell{\textbf{Thread}\\\textbf{Utilization}} & 
\makecell{\textbf{F}} \\
\hline\hline
SPHINCS$^+$-128f & 0.6875 & 0.6875 & 3 \\
\hline
SPHINCS$^+$-192f & 0.75   & 0.75   & 2 \\
\hline
\end{tabular}

\label{tab:fusion_utilization}
\end{table}

\subsubsection{Relax-FORS model}
Furthermore, considering the 256f parameter set, each FORS tree contains 512 leaf nodes, consuming up to 16 KB of shared memory per tree. Assigning one thread per leaf node allows at most two subtrees to run in parallel, while 256f requires processing 35 FORS trees in total. This leads to excessive synchronization and resource contention in terms of both threads and shared memory, thereby limiting warp occupancy.

\begin{figure}[h]
  \centering\includegraphics[width=1.0\linewidth]{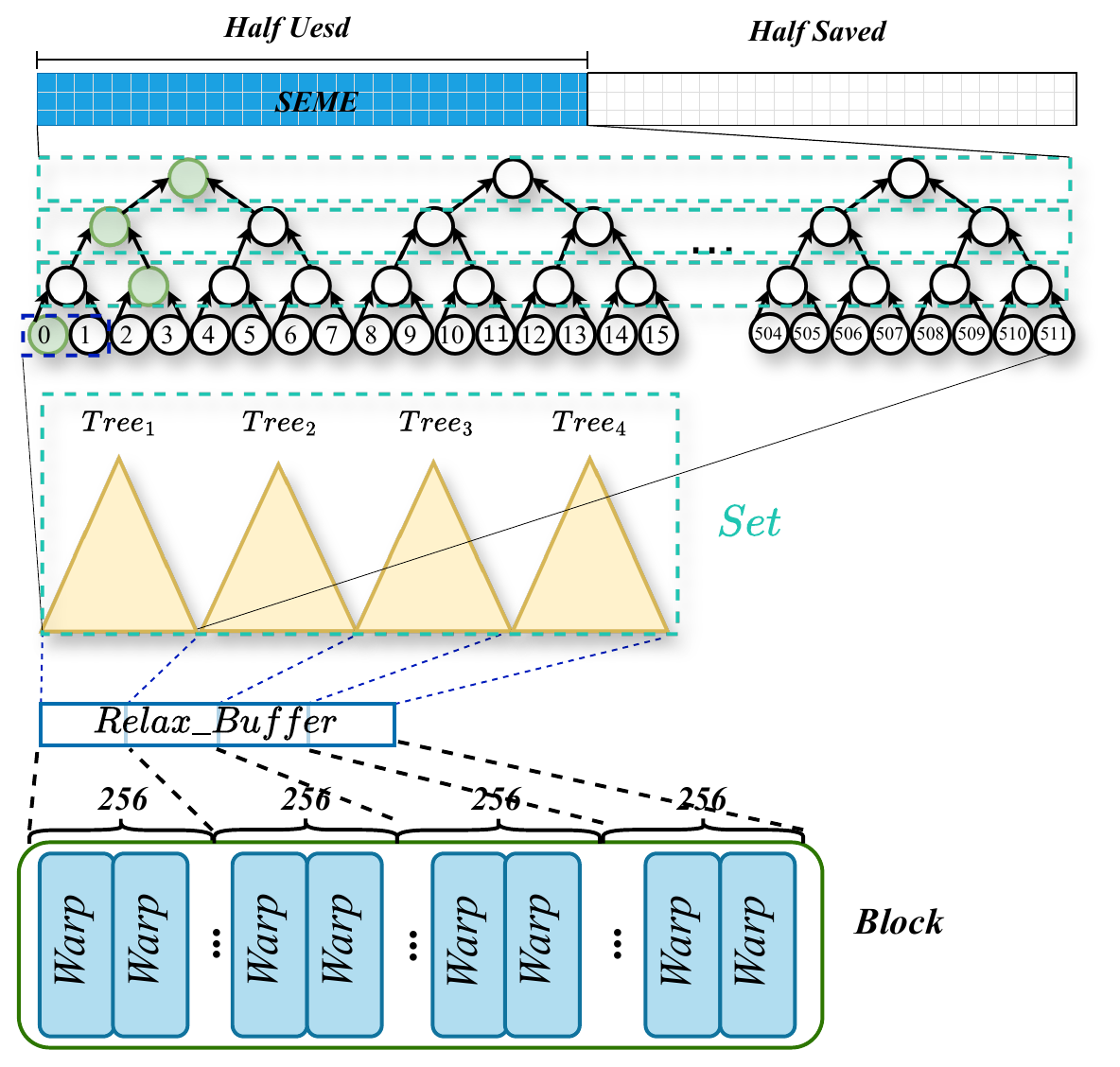}
  \caption{The Workflow of  \textit{Relax\_FORS}. The bottom layer of each $Set$ relies on a \textit{Relax\_buffer}: for each FORS tree, once a thread acquires it, it quickly computes the parent node and writes the parent node back to the shared memory of the upper layer. Thus, it reduces shared memory usage by half. The per-thread register constraint effectively reduces contention. In this $Set$, \textit{Relax\_buffer} reserves local space for eight leaf nodes to accommodate four trees.
}
  \label{fig:tree_relax}
\end{figure}

We propose \textit{Relax\_FORS} model to address these constraints. Observing that each tree’s shared-memory usage is halved when moving one level upward in the Merkle reduction (see Figure~\ref{fig:tree_reduction}), we introduce a \textit{Relax\_Buffer} at the bottom layer of each $Set$ and utilize complete shared memory for the upper layers. Instead of dedicating one thread per leaf node, a single thread is responsible for generating two leaf nodes and then storing them in the \textit{Relax\_Buffer}. This buffer is implemented in registers, whose per-thread organization aligns naturally with CUDA's fine-grained SIMT execution model. Each thread maintains a private register array, and we constrain its register usage to a threshold $R_t$ to reduce register pressure, avoid register spilling, and limit SM resource consumption. The overhead of this buffering strategy is negligible due to the low-latency register access. The \textit{Relax\_FORS} model shows that moderately reducing memory usage can improve performance by alleviating memory pressure and enhancing data locality. Figure~\ref{fig:tree_relax} provides an overview of the \textit{Relax\_FORS} workflow.

\subsection{Adaptive Selection of PTX branches}
\label{sec:PTX}

The hash-based structure of SPHINCS$^+$ inherently with intra-chain dependencies and inter-chain aggregation patterns, which pose a challenge for efficient thread allocation. HERO-Sign addresses this challenge by shifting the optimization of SHA2 computations to the compiler level. HERO-Sign leverages PTX-level instruction tuning to balance aggressive and conservative compiler optimizations. HERO-Sign further adopts an adaptive branch selection strategy tailored to each kernel’s behavior.

\subsubsection{The selection of PTX-level instructions}

The SHA-2 function processes each 512-bit message block through 64 rounds of bitwise logic, shift operations, and modular additions to produce a 256-bit digest. While these logical operations can be effectively optimized by the compiler, we observed that each round requires 16 big-endian data loads, traditionally implemented using multiple shift operations. We replace these shift-based conversions with the \texttt{prmt} instruction. While the \texttt{prmt} instruction has a higher latency than a simple \texttt{shl} instruction, it can replace multiple \texttt{shl} instructions with one single byte-level permutation, thus reducing \texttt{shl} register pressure.  \texttt{IADD3} instruction also has high usage rate in the SPHINCS$^+$-128f parameter set, we replaced it with the \texttt{mad} instruction, however, we observe that the compiler will still aggressively optimize the \texttt{mad} instruction into \texttt{IADD3} at SASS level. We introduce an auxiliary parameter \texttt{m} into the \texttt{mad} function, so the compiler will be misled to interpret the function as having four input parameters, which alters its optimization heuristics and will retain \texttt{mad} instructions at the SASS level shown in Figure~\ref {fig:ptx-snippets}.


\begin{figure}[!ht]
\centering
\begin{lstlisting}[style=mintedcpp]
// Permutation on a 32-bit value
asm volatile (
    "{\n\t"
    "  prmt.b32 %0, %1, 0, 0x0123;\n\t"
    "}\n"
    : "=r"(result)
    : "r"(input)
);

// Permutation on a 64-bit value
asm volatile (
    "{        \n\t"
    "  .reg .b32 x, y;              \n\t"
    "  mov.b64 {x, y}, %1;          \n\t"
    "  prmt.b32 x, x, 0, 0x0123;    \n\t"
    "  prmt.b32 y, y, 0, 0x0123;    \n\t"
    "  mov.b64 %0, {y, x};          \n"
    "}        \n"
    : "=l"(result)
    : "l"(input)
);

// MAD usage (example with m = 1)
asm volatile (
    "mad.lo.u32 %0, %1, %4, %2;\n\t"
    "mad.lo.u32 %0, %0, %4, %3;\n"
    : "=r"(result)
    : "r"(a), "r"(b), "r"(c), "r"(m)
);
\end{lstlisting}
\caption{
Examples of PTX inline assembly used for 32-bit and 64-bit permutation, as well as \texttt{mad.lo.u32} usage.  
All selected instructions operate on \texttt{uint32\_t} to maximize efficiency without compromising security.
}
\label{fig:ptx-snippets}
\end{figure}

\subsubsection{PTX Branch Selection at Kernel-Level}

Since the signature computation procedures vary across different kernels, and PTX does not always outperform native version due to restricted compiler optimization space, HERO-Sign considers PTX branch selection from a coarser-grained, kernel-centric perspective. Given the opaque and hardware-specific nature of the compilation pipeline (e.g., from \texttt{nvcc} or \texttt{ptxas} to different GPU architectures), quantitative analysis of PTX-level optimizations based solely on compiler rules remains infeasible. HERO-Sign adopts a more intuitive approach by providing both PTX and native branches, allowing profiling-based comparison to select the more performant configuration. The branch selection applied to different kernels are summarized in Table~\ref{tab:sphincs_ptx}. We also consider register-level factors based on the occupancy model in Equation~\ref{eq:reg_occupancy}. Due to the limited register file per SM on GPUs, increased register usage per thread leads to reduced warp occupancy.


\begin{equation}
\text{Occupancy} = \frac{1}{\mathcal{W}_{\max}} \cdot 
\left\lfloor 
\frac{\mathcal{R}_{\mathrm{total}}}{\mathcal{R}_{\mathrm{thread}} \times \mathcal{T}_{\mathrm{block}}} 
\right\rfloor 
\cdot \left( \frac{\mathcal{T}_{\mathrm{block}}}{32} \right)
\label{eq:reg_occupancy}
\end{equation}

\noindent
where $\mathcal{R}_{\mathrm{total}}$ is the total number of registers per SM, $\mathcal{R}_{\mathrm{thread}}$ is the number of registers allocated per thread, $\mathcal{T}_{\mathrm{block}}$ is the number of threads per block, and $\mathcal{W}_{\max}$ denotes the maximum number of warps supported per SM.

For instance, in  SPHINCS$^+$-256f, the baseline \texttt{TREE\_Sign} kernel requires 168 registers per thread, with a warp occupancy of only 19\%. By introducing the PTX branch, the register usage is reduced to 95 per thread, thereby improving occupancy to 37.5\%, a 1.97$\times$ increase compared to the native version. Furthermore, in \texttt{TREE\_Sign} kernel, generating a single leaf node with one thread involves 560, 816 and 1072 SHA-2 computations in SPHINCS$^+$-128f, SPHINCS$^+$-192f and SPHINCS$^+$-256f, respectively, the better performance of PTX in SPHINCS$^+$-256f suggests that PTX can help alleviate aggressive compiler optimizations.

\begin{table}[htbp]
  \centering
    \captionsetup{font=small} 
  \caption{PTX branch selection across signature kernels under different SPHINCS$^+$ parameter sets on RTX 4090 (Block Size = 1024). A \checkmark{} indicates that the PTX version outperformed native; \texttimes{} indicates the native version was retained.}
  \label{tab:sphincs_ptx}
  \renewcommand{\arraystretch}{1.0}  
  \setlength{\tabcolsep}{6pt}       
  \begin{tabular}{|l|c|c|c|}
    \hline
    \textbf{Parameter Set} & \textbf{FORS\_Sign} & \textbf{TREE\_Sign} & \textbf{WOTS+\_Sign} \\
    \hline\hline
    SPHINCS$^+$-128f & \checkmark & \texttimes & \texttimes \\
    SPHINCS$^+$-192f & \checkmark & \texttimes & \texttimes \\
    SPHINCS$^+$-256f & \checkmark & \checkmark & \checkmark \\
    \hline
  \end{tabular}
\end{table}


\subsubsection{Compilation-Level Branching Strategy}
Runtime branch selection inevitably results in multiple active code paths, which not only increases device memory usage but also imposes additional pressure on the instruction cache. HERO-Sign leverages \texttt{constexpr\ if} specialization to statically determine branch path at compile time, ensuring that each kernel includes only one single fixed execution path in the final binary. When all kernel selections resolve uniformly to either the native or PTX version, HERO-Sign opts to use a branch-free specialized copy, thereby eliminating unnecessary compilation overhead. Figure~\ref{fig:pt1} shows the implementation logic of compile-time branching.

\begin{figure}[t]
\centering
\begin{lstlisting}[style=mintedcpp]
// Define two selectable execution paths
template <bool UseOptimizedPath>
__device__ uint32_t SHA2(uint8_t* x) {
    if constexpr(UseOptimizedPath)
    { /* PTX Version */... }
    else{ /* Native */ ...}
}

// Each kernel forwards the compile-time policy to the SHA-2 routine. 

template <bool UseOptimizedPath>
__global__ void FORS_Sign(/* args */) {
    // ...
    (void)sha2_round<UseOptimizedPath>(/*x*/);
    // ...
}

template <bool UseOptimizedPath>
__global__ void TREE_Sign(/* args */) {
    // ...
    (void)sha2_round<UseOptimizedPath>(/*x*/);
    // ...
}

template <bool UseOptimizedPath>
__global__ void WOTS+_Sign(/* args */) {
    // ...
    (void)sha2_round<UseOptimizedPath>(/*x*/);
    // ...
}
// Predefine a bool variable to select between PTX and native paths.
dim3 grid(/*...*/), block(/*...*/);
// SPHINCS+_128f selection as follows
FORS_Sign<true><< <grid, block, ...>> >(...);
TREE_Sign<false><< <grid, block, ...>> >(...);
WOTS+_Sign<false><< <grid, block, ...>> >(...);
\end{lstlisting}
\caption{The Snippet illustrates the overview of the compile-time branch selection, from the kernel launch to the specific branches within the SHA-2 implementation.}
\label{fig:pt1}
\end{figure}

\subsection{Hybrid Memory Allocation}
\label{sec:HybridME}
HERO-Sign also exploits shared memory to accelerate internal array accesses by placing frequently accessed arrays in shared memory, effectively reducing the number of off-chip memory accesses. During FORS Fusion process, where memory access is more intensive, excessive usage of shared memory can reduce the number of active warps per SM, HERO-Sign mitigates shared memory pressure by allocating frequently used, read-only data, such as the initial state values, \texttt{sk\_seed}, \texttt{pk\_seed} and \texttt{state\_seed} to constant memory. Leveraging its broadcast capabilities, constant memory delivers access latency comparable to that of on-chip SRAM. In SPHINCS$^+$-192f, for the \texttt{Tree\_Sign} kernel, memory access is relatively infrequent, HERO-Sign allocates the corresponding read-only data directly in global memory and moderately leverages vectorized types such as \texttt{int4} and \texttt{int2} to enable coalesced memory accesses via \texttt{ldg.128} and \texttt{ldg.64} instructions.

\subsection{A Generalized Strategy for Eliminating Bank Conflicts}
\label{sec:free}
\begin{figure}[h]
  \centering\includegraphics[width=1.0\linewidth]{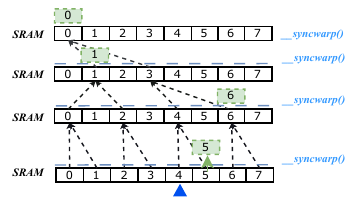}
  \caption{Tree-Based Reduction Process. \textcolor{customgreen}{\rule{0.6em}{0.6em}} highlights the signature nodes generated during the bottom-up process. \textcolor{blue}{\UParrow} represents the selected $leaf\_idx$, which guides the authentication path and signature nodes selection. Fast shared memory access enabled by on-chip SRAM facilitates frequent data access in Merkle Tree-based signature computation. By leveraging an even-odd access and storage pattern, shared memory can be utilized more efficiently without causing RAW data hazards and allow this computation to approach register-level performance \cite{10.1145/42190.42277}.}
  \label{fig:tree_reduction}
\end{figure}

 We utilize shared memory extensively during Tree-based Reduction process shown in Figure~\ref{fig:tree_reduction}. Bank conflicts in shared memory will occur when multiple threads within a warp access different addresses that reside in the same memory bank, causing serialized access and increased latency \cite{nvidia_cuda_guide}. Moreover, SPHINCS$^+$ defines three security levels, with each thread required to access 16 bytes, 24 bytes, and 32 bytes of data for Level 1, Level 2 and Level 3, respectively. Therefore, our bank conflict elimination strategy is designed to meet two key criteria:  (1) it must remain effective during Reduction process; and (2) it must be compatible with multi-bank, since each bank typically holds only 4 bytes.

\subsubsection{From Known Results to New Observations} 
\begin{figure}[h]
\centering\includegraphics[width=1.0\linewidth]{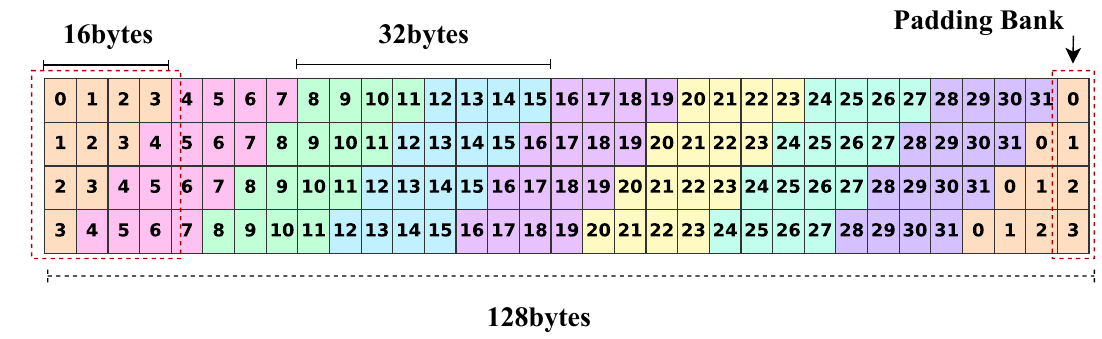}
  \caption{Each row represents a 128-byte transaction. The padding bank alters the access addresses of individual threads.}
  \label{fig: bk1}
\end{figure}

In CUDA architectures, a single shared memory transaction is typically aligned to 128 bytes. While bank conflicts often arise during a single shared memory transaction, they are not confined to accesses within a single warp~\cite{hitori2023cudaSharedMemory, nvidia2023bankconflict}. By inserting an extra bank after every 128 bytes, threads can be redirected to different memory banks, effectively eliminating bank conflicts, which much like conventional memory padding strategies. During Reduction process, where each thread $loads$ two nodes and $stores$ their parent node, since the applied padding alters the memory bank layout (see Figure~\ref{fig: bk1}), ensuring that all $load/store$ operations within a single transaction remain conflict-free. The padding strategy applies seamlessly to SPHINCS$^+$-128f and SPHINCS$^+$-256f, which involve 16-byte and 32-byte accesses per thread, respectively. We define the following padding formula: 

\begin{equation}
\underbrace{128}_{\text{A single memory transaction}} = \mathcal{B}_n \times 4 \times \mathcal{T}_h
\label{eq:smt}
\end{equation}
\noindent
where $\mathcal{B}_n$ denotes the number of banks each thread accesses, and $\mathcal{T}_h$ indicates the thread interval after which a padding bank is inserted.

\subsubsection{Extension to 24-Byte Bank Padding} 

\begin{figure}[h]
\centering\includegraphics[width=1.0\linewidth]{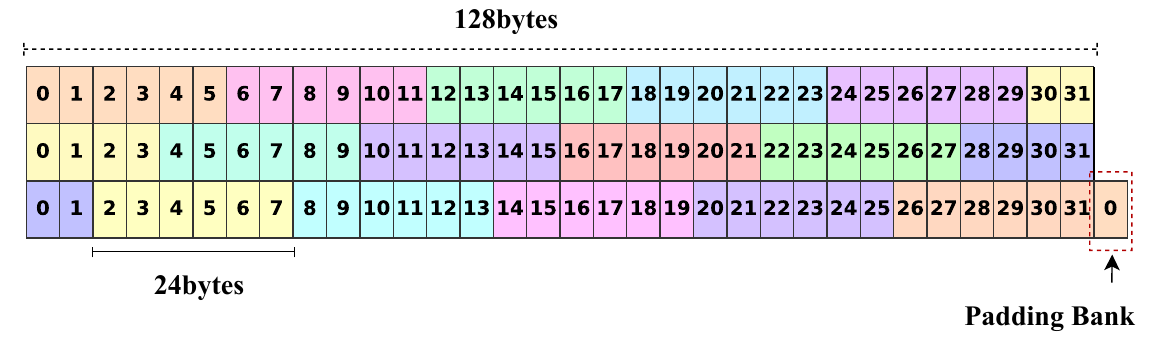}
  \caption{Address layout for 24-byte accesses. Illustration of cross-memory transaction redundancy in thread accesses, which persists until thread 16. Following Equation~\ref{eq:smt1}, a padding bank is inserted after the 16th thread.}
  \label{fig:bk2}
\end{figure}

In SPHINCS$^+$-192f, each thread access 24 bytes, which is not a multiple of 128; therefore, Equation~\ref{eq:smt} does not directly apply. Figure~\ref{fig:bk2} illustrates that 24-byte accesses break memory alignment, causing non-contiguous transactions, e.g., thread six access crosses a 128-byte boundary, and since its bank usage overlaps with thread one (banks 0–3), it is unclear whether new bank conflicts occur. We assume that the GPU hardware-level employs a coalescing strategy, allowing limited strided 128-byte accesses to be merged into a larger memory transaction. If this condition holds, the resulting access pattern within a warp induces only a 2-way bank conflict, which is more aligned with hardware-friendly scheduling strategies. Based on this, we derive the extended formula:

\begin{equation}
\underbrace{128 \times \mathcal{R}}_{\text{A single memory transaction region}} = \mathcal{B}_n \times 4 \times \mathcal{T}_h
\label{eq:smt1}
\end{equation}
\noindent
where $\mathcal{R}$ is a positive integer representing the number of contiguous 128-byte rows.


As shown in Table~\ref{tab:bank_conflict_padding}, our padding strategy reduces bank conflicts to near zero during Reduction process. This observation aligns with our initial hypothesis; to the best of our knowledge, this is the first time it has been observed that NVIDIA GPUs support a single shared memory transaction in multiples of 128 bytes.

\begin{table}[htbp]
\centering
  \captionsetup{font=small} 
\caption{Bank conflicts comparison between baseline and our padding strategy during Reduction process profiled by Nsight System.}
\label{tab:bank_conflict_padding}
\renewcommand{\arraystretch}{1.15}
\setlength{\tabcolsep}{4pt}
\resizebox{0.48\textwidth}{!}{%
\begin{tabular}{l|rr|cc|rr|cc}
\hline
\multirow{3}{*}{\textbf{Block = 1}} 
& \multicolumn{4}{c}{\textbf{FORS\_Sign}} 
& \multicolumn{4}{c}{\textbf{TREE\_Sign}} \\
\cline{2-9}
& \multicolumn{2}{c}{\textbf{Baseline}} & \multicolumn{2}{c}{\textbf{With Padding}} 
& \multicolumn{2}{c}{\textbf{Baseline}} & \multicolumn{2}{c}{\textbf{With Padding}} \\
\cline{2-9}
& \textbf{Load} & \textbf{Store} & \textbf{Load} & \textbf{Store} 
& \textbf{Load} & \textbf{Store} & \textbf{Load} & \textbf{Store} \\
\hline
\textbf{SPHINCS$^+$-128f} & 22,099,968 & 12,435,456 & 0 & 0 & 1568 & 704 & 1 & 0 \\
\textbf{SPHINCS$^+$-192f} & 64,152 & 30,096 & 0 & 0 & 1203 & 408 & 1 & 0 \\
\textbf{SPHINCS$^+$-256f} & 400,960 & 192,640 & 0 & 0 & 11,905 & 5,377 & 0 & 0 \\
\hline
\end{tabular}}
\end{table}

\subsection{Graph-based Multiple Batches of Message Signatures}
 Existing task-based graph construction frameworks, such as CUDA Graph~\cite{nvidia_cudagraph}, Intel oneAPI DPC++~\cite{intel_dpcpp}, StarPU~\cite{augonnet2011starpu}, MLIR Task DAG~\cite{mlir_taskdag}, and Taskflow~\cite{taskflow2023}, which enable efficient kernel packaging and concurrent execution. While traditional CUDA streams depend on host-initiated asynchronous submissions and GPU-side scheduling, it can lead to substantial kernel launch overhead when processing multiple batches. Task graph–based methods generally capture stream operations during execution and perform static analysis on kernels' dependencies and resource usage, effectively reducing kernel launch overhead at runtime. Once instantiated, the graph can be launched with a single call to execute multiple operations.

\begin{figure}[h]
  \centering\includegraphics[width=1.0\linewidth]{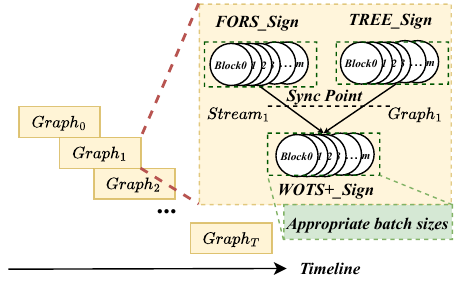}
  \caption{Block-based CUDA Graph construction strategies. \texttt{FORS\_Sign}, \texttt{TREE\_Sign} and \texttt{WOTS+\_Sign} kernels are organized into DAGs by adding explicit dependencies between nodes.}
  \label{fig:dag}
\end{figure}

\textbf{Graph Construction Strategies:} In analyzing kernel independence, only the signature generation of \texttt{WOTS\_Sign} depends on the root nodes generated by \texttt{FORS\_Sign} and \texttt{TREE\_Sign}, therefore, \texttt{FORS\_Sign} and \texttt{TREE\_Sign} can be scheduled earlier and execute concurrently. We assign one block to represent each message. We adopt CUDA Graph to construct the task graph, as it provides native CUDA support and enables hardware-level execution optimizations. For scenarios involving a large number of messages, we partition them into multiple batches and explore appropriate batch sizes for efficient overlapping (Figure~\ref{fig:dag}). For instance, in Block-based strategy, \( m \times T \) blocks are distributed across \( T \) CUDA graphs, each bound to a non-blocking stream, allowing concurrent execution and maximizing overlap across graphs. While orthogonal to traditional multi-streaming, this design benefits from CUDA Graph’s preallocation resource mechanism, enabling lower launch latency across multiple graph streams.


\section{Experimental Evaluation}
\label{sec:result}
\subsection{Implementation of HERO-Sign}
 HERO-Sign first employs an offline Tree Tuning algorithm to generate a candidate set of optimal FORS fusion configurations tailored to different GPU architectures. The most performance configuration is then selected based on profiling results. Leveraging the same profiling insights, HERO-Sign further selects the appropriate PTX or native implementation for each of the three major kernel types. Finally, the selected kernels are captured and assembled into a CUDA Graph for efficient graph-based execution.

\subsection{Evaluation Methodology}
\subsubsection{Baseline Selection} We select TCAS-SPHINCSp as our baseline for comparison, given that it is the fastest and the state-of-the-art (SOTA) publicly available GPU implementation of SPHINCS$^+$\cite{kim2024parallel}.

\subsubsection{Evaluation Metrics}
We define throughput, measured in kilo-signature operations per second (KOPS), as our primary performance metric. Kernel launch latency (in microseconds) is measured using Nsight Systems. We use Nsight Compute to obtain low-level profiling metrics, including warp occupancy, compute throughput and memory throughput ~\cite{nsight_compute}.

\subsubsection{Experimental Platform}

Table~\ref{tab:gpu_specs} summarizes our experimental platform configurations. We begin our evaluation on RTX 4090, where we present detailed performance results and optimization steps in Sections~\ref{sec:ps}, ~\ref{sec:op} and~\ref{sec:ss}. In Section~\ref{sec:s3}, we further extend HERO-Sign’s optimization strategies across multiple GPU architectures. To ensure a fair comparison, we maintain the same input length across all tests and compile each configuration using the appropriate \texttt{sm} version for the corresponding architecture.

\begin{table}[htbp]
\centering
\captionsetup{font=small} 
\caption{GPU architecture and corresponding CPU configurations. We use the PCIe interface for the following GPUs.}
\label{tab:gpu_specs}
\resizebox{1.0\linewidth}{!}{ 
\normalsize
\begin{tabular}{c|c|c|c|c}
\hline
\textbf{GPU}       & \textbf{Architecture} & \textbf{SM Version} & \textbf{Base Clock (MHz)} & \textbf{CPU (INTEL\textsuperscript{\textregistered} XEON\textsuperscript{\textregistered})} \\ 
\hline\hline
GTX 1070     & Pascal   & SM61  & 1506 & CPU E5-2620    \\
V100         & Volta    & SM70  & 1230 & PLATINUM 8575C \\
RTX 2080 Ti  & Turing   & SM75  & 1350 & CPU E5-2698    \\
A100         & Ampere   & SM80  & 1095 & GOLD 6330      \\
RTX 4090     & Ada      & SM89  & 2235 & GOLD 6530      \\
H100         & Hopper   & SM90  & 1035 & PLATINUM 8470  \\
\hline
\end{tabular}
}
\end{table}

\begin{figure*}[h]
\centering\includegraphics[width=1.0\linewidth]{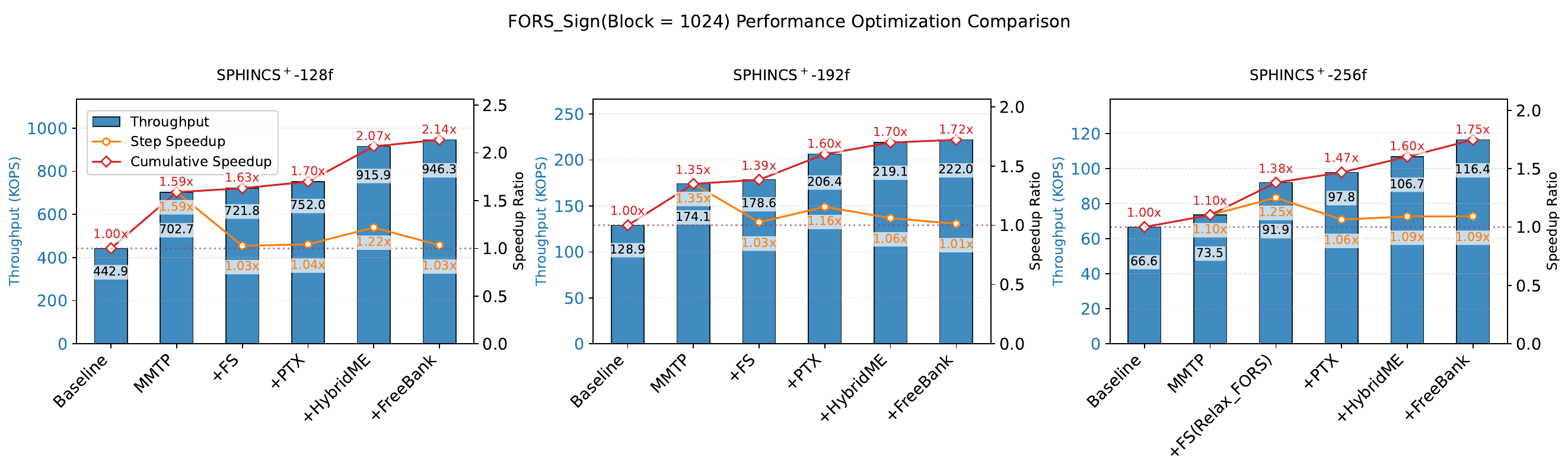}
  \caption{\texttt{FORS\_Sign} Performance Optimization Steps. Step Speedup refers to performance improvement introduced by each individual optimization over the previous stage, Cumulative Speedup measures the total acceleration compared to the baseline.}
  \label{fig:overall}
\end{figure*}

\begin{table*}[tp]
\centering
\fontsize{6.}{8}\selectfont
\caption{Kernel Performance comparison between Baseline and HERO-Sign (Block = 1024).}
\label{tab:performance_comparison}
\begin{tabular}{c|c|ccc|ccc|ccc|ccc}
\hline
\multirow{2}{*}{Parameter} & \multirow{2}{*}{Kernel} 
& \multicolumn{3}{c|}{Performance[KOPS]} 
& \multicolumn{3}{c|}{Occupancy [\%]} 
& \multicolumn{3}{c|}{Compute Throughput[\%]} 
& \multicolumn{3}{c}{Memory Throughput[\%]} \\
\cline{3-14}
& & Baseline & HERO-Sign & Speedup & Baseline & HERO-Sign & Imp & Baseline & HERO-Sign & Imp & Baseline & HERO-Sign & Imp \\
\hline\hline
\multirow{3}{*}{SPHINCS$^+$-128f} 
& \texttt{FORS\_Sign} & 442.9 & 946.3 & 2.14× & 27.09 & 36.02 & 1.33× & 45.18 & 56.37 & 1.25× & 11.26 & 9.83 & 0.87× \\
& \texttt{TREE\_Sign} & 125.2 & 157.7 & 1.26× & 23.65 & 23.88 & 1.01× & 92.87 & 97.67 & 1.05× & 247.0 & 1.88 & 0.76× \\
& \texttt{WOTS+\_Sign} & 2493.1 & 4915.7 & 1.97× & 42.36 & 46.54 & 1.10× & 43.63 & 34.55 & 0.79× & 73.70 & 69.94 & 0.95× \\
\hline
\multirow{3}{*}{SPHINCS$^+$-192f} 
& \texttt{FORS\_Sign} & 128.9 & 222.0 & 1.72× & 32.74 & 47.05 & 1.44× & 44.69 & 54.48 & 1.22× & 10.21 & 8.26 & 0.81× \\
& \texttt{TREE\_Sign} & 88.2 & 93.6 & 1.06× & 23.83 & 23.87 & 1.00× & 95.57 & 97.76 & 1.02× & 4.73 & 2.54 & 0.54× \\
& \texttt{WOTS+\_Sign} & 1457.6 & 2464.9 & 1.69× & 31.44 & 35.09 & 1.12× & 24.50 & 22.37 & 0.91× & 82.49 & 84.23 & 1.02× \\
\hline
\multirow{3}{*}{SPHINCS$^+$-256f} 
& \texttt{FORS\_Sign} & 66.6 & 116.4 & 1.75× & 32.60 & 63.76 & 1.96× & 42.42 & 66.37 & 1.56× & 20.71 & 13.55 & 0.65× \\
& \texttt{TREE\_Sign} & 36.4 & 44.9 & 1.23× & 18.53 & 62.43 & 3.37× & 72.38 & 96.17 & 1.33× & 5.46 & 10.42 & 1.91× \\
& \texttt{WOTS+\_Sign} & 776.8 & 1570.9 & 2.02× & 35.37 & 35.47 & 1.00× & 11.93 & 12.77 & 1.07× & 88.19 & 86.80 & 0.98× \\
\hline
\end{tabular}
\end{table*}

\subsection{Performance Steps of \texttt{FORS\_Sign}}\
\label{sec:ps}
Figure~\ref{fig:overall} illustrates the optimization steps of \texttt{FORS\_Sign} kernel. We begin by enabling parallel construction of multiple Merkle trees, \textit{MMTP} (Section~\ref{sec:MMTP}). We then apply a FORS fusion strategy, \textit{+FS} (Section~\ref{sec:FS}). Next, we apply PTX-level instruction tuning, \textit{+PTX} (Section~\ref{sec:PTX}). To mitigate shared memory contention, we introduce a hybrid memory allocation scheme, \textit{+HybridME} (Section~\ref{sec:HybridME}). Finally, we eliminate shared memory bank conflicts through a lightweight bank padding technique, \textit{+FreeBank} (Section~\ref{sec:free}). For both \texttt{128f} and \texttt{192f}, the most effective performance improvement comes from \textit{MMTP}, which fully leverages the inherent independence across Merkle tree computations to improve CUDA Core utilization. In case of \texttt{256f}, we further enhance its parallelism by introducing the \textit{Relax\_FORS} model, resulting in a 1.38$\times$ speedup over Baseline. \textit{+FreeBank} offers limited benefit, as it mainly helps workloads sensitive to shared memory bank conflicts. For \texttt{192f}, eliminating 24-byte bank conflicts yields only a 1\% speedup. We hypothesize that such inefficiency may stem from limited native instruction support for aligned to more than 128-byte boundary memory transactions on modern GPU architectures. We also observe that different optimization strategies yield varying benefits across parameter sets. For instance, \textit{+HybridME} achieves the most pronounced step speedup of 1.22$\times$ under \texttt{128f}. This is attributed to a higher degree of FORS tree fusion, where each thread frequently accesses overlapping memory regions. By leveraging constant memory broadcasting, \textit{HybridME} effectively mitigates the pressure from intensive memory reads.

\subsection{Overall Kernel Performance Comparison of SPHINCS$^+$}

\label{sec:op}
\renewcommand{\arraystretch}{1.3}

\begin{figure*}[h]
\centering\includegraphics[width=1.\linewidth]{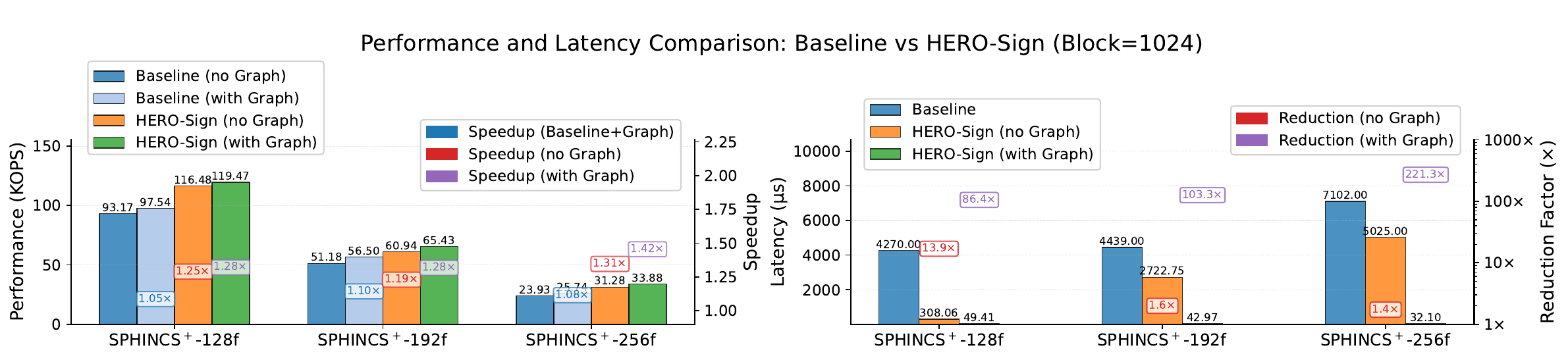}
  \caption{Performance (KOPS) and Kernel Launch Latency (µs) comparison between the Baseline and the fully optimized HERO-Sign, with and without CUDA Graph integration (Graph instantiation time is excluded).}
  \label{fig:all}
\end{figure*}

When integrated, our final performance improvements yield up to 2.14$\times$, 1.26$\times$ and 2.02$\times$ speedups in \texttt{FORS\_Sign}, \texttt{TREE\_Sign} and \texttt{WOTS+\_Sign}, respectively, across the \texttt{128f}, \texttt{192f} and \texttt{256f} parameter sets shown in Table~\ref{tab:performance_comparison}. Benefiting from \textit{+PTX} instruction tuning, the \texttt{TREE\_Sign} kernel under \texttt{256f} achieves a 3.37$\times$ increase in warp occupancy. In contrast, we observe a reduction in compute throughput for \texttt{WOTS+\_Sign} under \texttt{128f} and \texttt{192f}. This is expected, as our optimization rewrites expensive division and modulo operations into more efficient bitwise shifts and \texttt{\&} operations. Additionally, we observe a decrease in memory throughput across several kernels, with the maximum reduction reaching 0.54$\times$. This is primarily due to \textit{HybridME} effectively reducing off-chip memory traffic. We also observe a 1.91$\times$ improvement in memory throughput for \texttt{TREE\_Sign} under the \texttt{256f} configuration, where shared memory bank conflicts are particularly pronounced. This improvement stems from our 32-byte bank conflict elimination strategy, which effectively enhances the efficiency of shared memory transactions. We compare the overall performance of all kernels under three configurations(Figure~\ref{fig:all}): Baseline, HERO-Sign without CUDA Graph and HERO-Sign with CUDA Graph. HERO-Sign with CUDA Graph consistently achieves the best performance, demonstrating that task graph construction can significantly reduce inter-kernel idle time. In \texttt{128f}, idle time is largely mitigated through constraining register allocation. However, in \texttt{192f} and \texttt{256f}, resource contention among kernels introduces idle periods that cannot be fully resolved in software. By packaging multiple kernels into a single execution graph relying on the CUDA driver, CUDA Graph effectively addresses this issue and reduces kernel launch latency by up to two orders of magnitude (221.3$\times$).

\textbf{Comparison with FPGA and ASIC Implementations shown in Table ~\ref{tab:cross_platform_boxed}.}
Quentin Berthet \textit{et al.} focus on minimizing resource utilization on \textit{Xilinx XZU3EG} devices\cite{berthet2021area}. 
In contrast, HERO-Sign achieves up to \textbf{$10^3\times$} higher throughput, while also demonstrating substantial energy efficiency improvements. 
Specifically, HERO-Sign reduces the PPS (per-signature power consumption) by 133$\times$ and 158$\times$ compared to their implementation. 
Similarly, Amiet \textit{et al.} optimize SPHINCS$^+$ on the \textit{FPGA Artix-7} platform using the \textit{SHAKE} instance\cite{amiet2018fpga}. 
Despite this difference in the hash function, HERO-Sign still achieves significant throughput improvements of 120.68$\times$, 76.98$\times$ and 84.70$\times$ for the \texttt{128f}, \texttt{192f} and \texttt{256f} parameter sets, respectively, 
while maintaining notably lower PPS. We further include the limited available ASIC-optimized implementation, \textit{SPHINCSLET}, for reference\cite{10.1145/3728469}. 
Compared to this design, HERO-Sign achieves remarkable throughput improvements of 
229.75$\times$, 327.15$\times$ and 338.8$\times$, respectively, 
highlighting the substantial efficiency advantage of our GPU-based architecture over existing hardware counterparts.

\begin{table}[t]
\centering
\caption{Cross-Platform Comparison of SPHINCS$^+$ Variants (Throughput(KOPS), PPS(Watt)). PPS: Power Consumption Per signature.}
\label{tab:cross_platform_boxed}
\resizebox{\linewidth}{!}{
\begin{tabular}{|l|c|c|c|c|}
\hline
\textbf{Variant} & \textbf{GPU (RTX4090)} & \textbf{FPGA} & \textbf{FPGA} & \textbf{ASIC} \\ 
\hline
\textbf{HASH Algorithm} & SHA256 & SHA256 & SHAKE256 & SHA256 \\
\hline\hline
Throughput (KOPS) & HERO-Sign & Berthet et al.~\cite{berthet2021area} & Amiet et al.~\cite{9217834} & SPHINCSLET~\cite{10.1145/3728469} \\
\hline
SPHINCS+-128f & 119.47 & 0.016 & 0.99 & 0.52 \\
128f PPS (Watt) & 0.003 & 0.4 & 9.76 & No Support \\
\hline
SPHINCS+-192f & 65.43 & No Support & 0.85 & 0.20 \\
192f PPS (Watt) & 0.002 & No Support & 9.69 & No Support \\
\hline
SPHINCS+-256f & 33.88 & 0.00057 & 0.40 & 0.10 \\
256f PPS (Watt) & 0.003 & 0.474 & 9.80 & No Support \\
\hline
\end{tabular}
}
\end{table}

\begin{table}[t]
\centering
\caption{CPU AVX2 Performance Comparison (KOPS)\cite{alter2021optimizing}}
\label{tab:cpu_avx2_box}
\resizebox{1.0\linewidth}{!}{
\begin{tabular}{|l|c|c|c|}
\hline
\textbf{Implementation} & \textbf{SPHINCS$+$-128f} & \textbf{SPHINCS$+$-192f} & \textbf{SPHINCS$+$-256f} \\
\hline\hline
AVX2 (Single Thread) & 0.143 & 0.087 & 0.044 \\
\hline
AVX2 (16 Threads) & 0.828 & 0.560 & 0.356 \\
\hline
\end{tabular}
}
\end{table}

\textbf{Comparison with AVX2 Implementations.}
Table~\ref{tab:cpu_avx2_box} presents the performance results of the AVX2-based CPU implementations\cite{alter2021optimizing}. 
When compared to the multi-threaded AVX2 baseline, HERO-Sign achieves substantial throughput improvements of 
144.29$\times$, 116.84$\times$ and 95.17$\times$ 
for the \texttt{128f}, \texttt{192f} and \texttt{256f} parameter sets, respectively.

\subsection{Sensitivity Study}
\label{sec:ss}

\subsubsection{Sensitivity on Block Sizes \& Batch Sizes}
\label{sec:s1}

\begin{figure}[h]
\centering\includegraphics[width=1\linewidth]{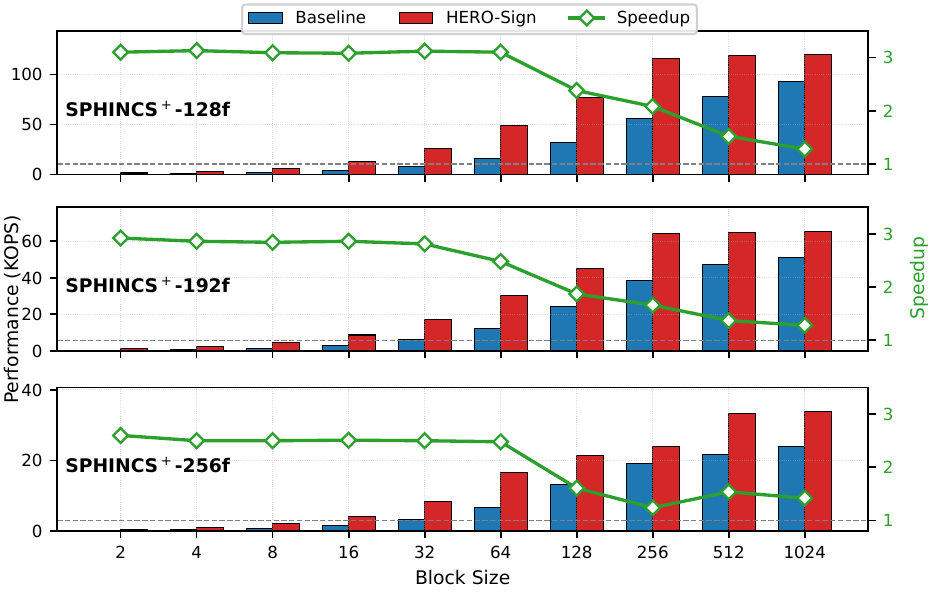}
  \caption{Performance (KOPS) comparison between Baseline and HERO-Sign (with Graph) under varying Block Sizes.}
  \label{fig:allblock}
\end{figure}

We evaluate the performance impact of varying block size from 2 to 1024(Figure~\ref{fig:allblock}). HERO-Sign maintains consistently high speedups across all three parameter sets within the 2–64 block size range, achieving 3.10$\times$–3.10$\times$ for \texttt{128f}, 2.92$\times$–2.48$\times$ for \texttt{192f}, and 2.60$\times$–2.48$\times$ for \texttt{256f}. As the block size increases, speedup begins to degrade due to approaching the GPU’s resource limits. For \texttt{256f}, the speedup drops most noticeably at block size 128, as the introduction of the \textit{Relax\_Buffer} intensifies thread contention. However, at block size 512, speedup improves, reflecting the effectiveness of \textit{Relax\_Buffer} in alleviating shared memory pressure. Therefore, for RTX 4090, when considering maximizing throughput, larger batch sizes (e.g., $\geq$512) are preferred unless PCIe transfer becomes the bottleneck, to enable better overlap between host-device data transfers and computation, a smaller batch size near 64 is optimal.

\subsubsection{Sensitivity on Compilation overhead}

As discussed earlier, we introduce compile-time branching to enable automatic selection between PTX and native paths. To evaluate its impact, we measure the average compilation time across batch sizes from 2 to 1024 for all three parameter sets in Table~\ref{tab:avg-compile-time}. Interestingly, our approach reduces compilation time by 1.28$\times$, 1.07$\times$ and 1.26$\times$ compared to the Baseline. In other words, the reduction in compilation time from limiting the compiler’s optimization space via PTX outweighs the overhead introduced by template instantiation at the compilation stage.

\label{sec:s2}
\begin{table}[t]
\centering
\captionsetup{font=small} 
\caption{Average compilation time (in seconds) from Block Size 2 to 1024 for Baseline and HERO-Sign.}
\label{tab:avg-compile-time}
\begin{tabular}{|l|c|c|c|}
\hline
\textbf{Parameter} & \textbf{Baseline} & \textbf{HERO-Sign} & \textbf{Speedup} \\
\hline\hline
SPHINCS$+$-128f & 18.68 & 14.61 & 1.28$\times$ \\
\hline
SPHINCS$+$-192f & 23.25 & 21.72 & 1.07$\times$ \\
\hline
SPHINCS$+$-256f & 24.19 & 19.18 & 1.26$\times$ \\
\hline
\end{tabular}
\end{table}
\subsubsection{Sensitivity on Input Sizes} We also evaluate performance across input lengths of 1K, 2K, 3K and 4K, with the block size fixed at 1024. HERO-Sign achieves average speedups of 1.30$\times$, 1.28$\times$ and 1.45$\times$ on \texttt{128f}, \texttt{192f} and \texttt{256f}, respectively. In practice, input messages of varying lengths are first hashed to produce a digest, from which the \texttt{leaf\_idx} values are derived. These indices determine the signature path in SPHINCS$^+$. However, since the tree structure and number of signing operations are fixed, the overall computational workload remains constant regardless of the input length.

\subsection{Extension to different GPU Architectures}
\label{sec:s3}
We extend HERO-Sign to a range of GPU architectures and apply the offline Tree Tuning algorithm to determine the maximum configurable dynamic shared memory per block on each platform. This allows us to generate a candidate set of optimized FORS fusion strategies tailored to the architectural constraints, from which the best-performing configuration is selected. To mitigate performance degradation caused by control-flow divergence in irregular fusion configurations, we prefer fusion strategies that minimize conditional branches. 

\begin{figure}[h]
\centering\includegraphics[width=1\linewidth]{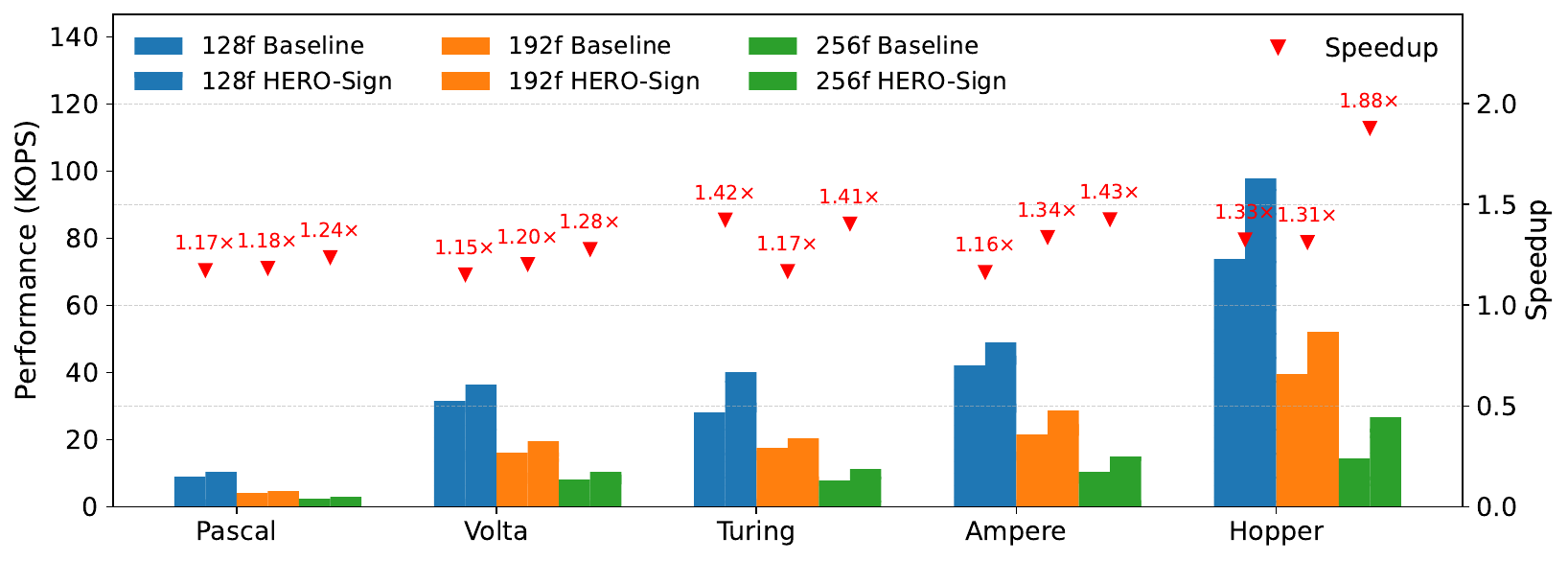}
  \caption{Performance Comparison: Baseline vs HERO-Sign(with Graph) (Block=1024) Across GPU Architectures.}
  \label{fig:gpuall}
\end{figure}

Figure~\ref{fig:gpuall} presents the performance comparison across architectures. The Pascal architecture exhibits the lowest throughput and speedup due to its limited number of CUDA cores (1920) and significantly smaller shared memory per SM (64~KB), which restricts warp-level concurrency and fusion depth. In contrast, the Hopper architecture provides up to 228~KB of shared memory per SM, which allows for a larger number of active warps to be resident concurrently. Leveraging this, HERO-Sign achieves its highest performance gain on \texttt{256f}, with a speedup of 1.88$\times$. However, across all architectures, RTX~4090 consistently delivers the highest absolute performance. For instance, on \texttt{256f}, HERO-Sign achieves 33.88 KOPS on RTX~4090 versus 26.63 KOPS on H100. Although H100 has slightly more CUDA cores (16,896 vs. 16,384), its base clock frequency is significantly lower (1,035~MHz vs. 2,235~MHz on RTX~4090). Given that SPHINCS$^+$ is an arithmetic instruction-bound workload, performance scales closely with core frequency. Assuming ideal IPC (Instructions Per Cycle), the instruction throughput can be approximated as:
\[
\text{Throughput} \propto \text{Core Count} \times \text{Frequency}
\]

Thus, despite a smaller core count, RTX~4090 achieves a higher instruction throughput due to its 2.16$\times$ frequency advantage, which aligns with the observed performance superiority.

\vspace{-0.3cm}
\section{Conclusion}
\label{sec:con}
In this paper, we propose HERO-Sign, a GPU-based accelerated framework for SPHINCS$^+$ signature generation. HERO-Sign achieves significant speedups across multiple GPU architectures. Maximizing SPHINCS$^+$ throughput remains a critical step toward its practical deployment as a post-quantum cryptography scheme, and our work serves as a foundation for continued exploration in this direction.

\section*{Acknowledgment}
This work is supported by the United States National Science Foundation under Grant No. 2530705.




\bibliographystyle{IEEEtranS}
\bibliography{refs}
\end{document}